\documentclass[twocolumn,eqsecnum,superscriptaddress,showpacs]{revtex4}
\usepackage{amsmath,amssymb}
\usepackage{bm}

\begin{document}
\title{Physics of Renormalization Group Equation in QED}
\author{Takehisa \textsc{Fujita}} \email{fffujita@phys.cst.nihon-u.ac.jp}
\affiliation{Department of Physics, Faculty of Science and Technology
Nihon University, Tokyo, Japan }%

\date{\today}

\begin{abstract}

It is shown that the renormalization group (RG) equation in QED can only describe 
the finite size effects of the system. The RG equation is originated 
from the response of the renormalized coupling constant for the change of the system 
size $L$. The application of the RG equation to the continuum limit treatment 
of the lattice gauge theory, therefore, does not make sense, and the 
well-known unphysical result of the lattice gauge theory with Wilson's action  
cannot be remedied any more.

\end{abstract}

\pacs{11.10.Gh,12.38.Gc,11.15.Ha}
\maketitle

\section{ Introduction}

The concept of renormalization is originated from the perturbative treatment 
in quantum field theory. In QED, one cannot calculate physical observables  
in a non-perturbative fashion, and therefore, one should employ the perturbation 
theory. In QED, one can well construct the free fermion 
and free gauge field Fock spaces (free QED space) which can be characterized by 
the number of degrees of freedom $N$ and the system size $L$ 
$$ {\rm Free \ \ QED  \ \ space } \ : \   (L,  \Lambda) \ \ {\rm with} \ \  
\Lambda={2\pi \over L}N . $$
Therefore, one can develop the perturbation 
theory in which one treats perturbatively the interaction term which is gauge invariant 
together with the fermion current conservation. In this case, all 
the physical quantities must be described in terms of the free QED space terminology.  

In the perturbation theory, the self-energy diagrams become divergent, and therefore 
one should employ the renormalization scheme which is very successful in QED. 
All of the infinities arising from the self-energy diagrams can be well renormalized 
into the redefinition of the fermion mass, coupling constant and wave functions. 
In addition, finite contributions are controlled and evaluated precisely, 
and some of them are compared with experiments, and one finds that the renormalization 
scheme is all consistent with experiments \cite{bd,ryder}. 

After the renormalization, one realizes that the renormalized charge $e$ should 
depend on the global quantity  $L$ which characterizes 
the unperturbed QED space, that is 
$$ e=e(L) . $$
This is reasonable since one evaluates the renormalization constant in the free QED 
space terminology, and therefore calculated quantities should depend 
on the properties of the unperturbed QED space in some way or the other. 
If one makes this dependence into the differential equation, then it becomes 
the RG equation which can give the description of the finite size effects of QED. 

This must be the basic story of the RG equation which should be given by old field 
theory experts with the fundamental renormalization scheme of QED \cite{fujita}. 

Now, the understanding of the renormalization group equation in recent years is quite 
different and somewhat puzzling. This difference should be originated from the dimensional 
regularization scheme \cite{thv1,thv2} even though the dimensional regularization itself 
has no problem in the renormalization procedure. Indeed, it can give just the same 
renormalization scheme as the cutoff momentum method, and furthermore it has some advantage 
since it is simpler and the calculation can be carried out in a covariant way. 

However, when one wishes to derive the RG equation, then it becomes 
problematic. In the dimensional regularization, one introduces a new scale $\lambda_s$ 
when one evaluates the momentum integral
$$ \int{d^4k\over (2\pi)^4} \rightarrow \lambda_s^{4-D}\int{d^Dk\over (2\pi)^D} 
\eqno{(1.1)} $$
where $\lambda_s$ is a parameter which has a mass dimension in order to 
compensate the unbalance of the momentum integral dimension. 
$D$ is set to be 
$$ D=4-\epsilon  \eqno{(1.2)}  $$
where $\epsilon$ is an infinitesimally small constant. In this dimensional regularization,  
the momentum integral is cut out by the four dimensional Euclidean space which is 
a compact space. This is in contrast to the normal way of regularization with the cutoff 
$\Lambda$ confined in the box $V=L^3$. 

What is $\lambda_s$ ? As long as one employs the perturbation theory, one has to 
find out a corresponding quantity of $\lambda_s$ in free QED  space, and otherwise one 
would discover new physics without doing any physics ! 
The corresponding quantity of $\lambda_s$ in free QED  space must be a global 
quantity, and the only possible candidate should be 
$$ \lambda_s \rightarrow {1\over L} .  \eqno{(1.3)}  $$
In fact, one may find some correspondence between the two 
different regularizations, for example
$$ \Lambda \sim \lambda_s \exp\left({1\over \epsilon}\right), \ \ \ \ \ 
L\sim {1\over \lambda_s}, 
\ \ \ \ \ N \sim \exp\left({1\over \epsilon}\right). \eqno{(1.4)}  $$
In this way, one can construct the dimensional regularization scheme in terms of 
the original free QED  space terminology. 

Therefore, it is clear that one should not apply the RG equation to the treatment 
of the continuum limit in the lattice gauge theory \cite{wil,rothe,asaga}. 
In the path integral formulation, the coupling constant $e$ is just a constant like 
the bare charge since it is a non-perturbative treatment, and therefore unphysical results 
which are obtained by Wilson cannot be remedied by any means.

\section{Renormalization scheme}
Before coming to the renormalization group equations, we should review the renormalization 
procedure in QED so as to clarify where the problem comes about. 

\subsection{Free QED space}
First, we start from the QED Lagrangian density $\cal L$ which is composed 
of the unperturbed Lagrangian density ${\cal L}_0 $ and the interaction term 
${\cal L}_I $
$$ {\cal L}_0= \bar\psi (p \llap/ -m) \psi -{1\over 4}F_{\mu \nu} F^{\mu \nu} 
\eqno{(2.1)} $$
$$ {\cal L}_I = - e A^\mu \bar\psi \gamma_\mu \psi . \hspace{2cm} \eqno{(2.2)}  $$
In this case, the unperturbed Hamiltonian $H_0$ can be constructed 
from the Lagrangian density ${\cal L}_0$.  
The Hilbert space of the quantized  Hamiltonian $\hat{H}_0$ can be well constructed 
since one finds the exact eigenvalues and eigenstates of the $\hat{H}_0$. 
In this case, the QED space can be specified by  the box length $L$ and the cutoff 
momentum $\Lambda$ as well as by the energies and momenta of the free fermion and 
free gauge field states
$$\underline{ {\rm Free \ \ QED  \ \ space} \ \ {\rm with} \ \ \  (L, \Lambda) } 
\hspace{1cm} $$
$$ {\rm Fermions}: \ \ E_p= \pm \sqrt{\bm{p}_n^2+m^2}, 
\ \ \bm{p}_n={2\pi \bm{n}\over L} \hspace{0.5cm} \eqno{(2.3a)}  $$
$$ {\rm Gauge \  fields}: \ \ \  \omega_k=|\bm{k}_n|, \ \  \ \bm{k}_n={2\pi \bm{n}\over L} 
\hspace{1cm} \eqno{(2.3b)}  $$
where $n_i$ runs as
$$ \ n_i=0, \pm 1, \cdots, \pm N 
\ \  {\rm with} \ \  \Lambda={2\pi N\over L} . \eqno{(2.3c)} $$
The maximum number of freedom $N$ is taken to be the same between the fermion 
and the gauge fields. 
The perturbative evaluation can be made within this Hilbert space, and one can 
calculate physical quantities in terms of the expansion of the coupling constant $e$. 
In other words, all the physical observables should be expressed in terms of the free 
QED space. This simple but important fact has been overlooked in deriving 
the renormalization group equation.

\subsection{Mass Renormalization} 

The fermion self-energy $\Sigma(p)$ can be evaluated with the dimensional 
regularization as
$$ \Sigma(p)=-ie^2\lambda_s^{4-D}\int {d^Dk\over(2\pi)^D}
\gamma_\mu { 1\over p \llap/-k \llap/-m }\gamma^\mu { 1\over k^2 } $$
$$={e^2\over 8\pi^2\epsilon}(-p \llap/ +4m)+ \textrm{finite terms}  \hspace{0.1cm} 
 \eqno{(2.4)}  $$
Therefore, the Lagrangian density of the free fermion part 
$$ {\cal L}_{\textrm {F}}=\bar\psi (p \llap/ -m) \psi \eqno{(2.5)}  $$
should be modified, up to one loop contributions, by the counter term 
$ \delta{\cal L}_{\textrm {F}}$
$$ \delta{\cal L}_{\textrm {F}}=  \bar\psi\left[{e^2\over 8\pi^2\epsilon} 
(-p \llap/ +4m) \right] \psi .  \eqno{(2.6)}   $$
In this case, the total Lagrangian density of fermion becomes
$$ {\cal L'}_{\textrm {F}}=\bar\psi_{\rm b} (p \llap/ -m_0) \psi_{\rm b}
+ \textrm{finite terms}  \eqno{(2.7)}  $$ 
where one introduces the wave function renormalization and the bare mass $m_0$
$$ \psi_{\rm b} \equiv \sqrt{Z_2} \psi   \eqno{(2.8a)}  $$
$$ m_0 =m \left(1+{e^2\over 8\pi^2\epsilon}\right) 
\left(1-{e^2\over 2\pi^2\epsilon}\right) \simeq m
-{3me^2\over 8\pi^2\epsilon}   \eqno{(2.8b)}  $$
where one should always keep up to order of $e^2$. 
Here, one defines $Z_2$ as
$$ Z_2 =1-{e^2\over 8\pi^2\epsilon}=1-{e^2\over 8\pi^2} 
\ln \left({\Lambda\over m}\right) \eqno{(2.9)} $$
where we also show the calculation of the cutoff momentum scheme. 
Therefore, the total Lagrangian density 
has just the same shape as the original one, and thus it is renormalizable. 

\subsection{Vacuum Polarization}

The divergent contributions to the self-energy of photon can be described 
in terms of the vacuum polarization
$$ \Pi^{\mu \nu}(k)=i\lambda_s^{4-D}e^2\int {d^Dp\over(2\pi)^D} 
Tr \left[ \gamma^\mu {1\over p \llap/-m  } \gamma^\nu 
{1\over p \llap/-k \llap/-m  }\right] $$
$$ ={e^2\over 6\pi^2\epsilon}(k^\mu k^\nu-g^{\mu \nu}k^2)+ \textrm{finite terms}.  
\hspace{0.1cm} \eqno{(2.10)}  $$
Defining $Z_3$ and the vector field renormalization by
$$ Z_3 =1-{e^2\over 6\pi^2\epsilon}
=1-{e^2\over 6\pi^2}\ln \left({\Lambda\over m}\right)    \eqno{(2.11)}  $$
$$ A_{\rm b}^\mu \equiv \sqrt{Z_3} A^\mu  \hspace{3.0cm}   \eqno{(2.12)}  $$
one can rewrite the Lagrangian density of the gauge field as
$$ {\cal L'}_{\textrm {GF}}= -{Z_3\over 4}F_{\mu \nu} F^{\mu \nu}= 
-{1\over 4} \left( \partial^\mu A_{\rm b}^{\nu} -\partial^\nu A_{\rm b}^{\mu} 
 \right)^2 +\cdots .  \eqno{(2.13)}  $$

\subsection{Vertex Corrections}
The vertex corrections can be evaluated as
$$ \Lambda^\mu (p',p)= -i\lambda_s^{4-D}e^2\int {d^Dk\over(2\pi)^D} \hspace{3cm}   $$
$$ \times \left[  \gamma^\nu {1\over {p \llap/}'-k \llap/-m  } \gamma^\mu 
{1\over p \llap/-k \llap/-m  } \gamma_\nu {1 \over k^2} \right] $$
$$ ={e^2\over 8\pi^2\epsilon} \gamma^\mu + \textrm{finite terms} . \hspace{1.5cm} 
\eqno{(2.14)}  $$
Therefore, the counter term of the interaction Lagrangian density $ \delta{{\cal L}}_I $ 
becomes
$$ \delta{{\cal L}}_I=e\lambda_s^{\epsilon\over 2} \left({e^2\over 8\pi^2\epsilon}\right) 
A^\mu \bar\psi \gamma_\mu \psi. \eqno{(2.15)}   $$
In this case, the total interaction Lagrangian density can be written as
$$ {{\cal L}'}_I  = -Z_1e\lambda_s^{\epsilon\over 2} 
A^\mu \bar\psi \gamma_\mu \psi +\textrm{finite terms} \eqno{(2.16)}  $$
where $Z_1$ is defined as
$$ Z_1 \equiv 1-{e^2\over 8\pi^2\epsilon} 
= 1-{e^2\over 8\pi^2}\ln \left({\Lambda\over m}\right). \eqno{(2.17)} $$
The interaction Lagrangian density can be rewritten 
in terms of the bare quantities 
$$ \psi_{\rm b} \equiv  \sqrt{Z_2} \psi, \ \ \ \  A_{\rm b}^\mu \equiv \sqrt{Z_3} A^\mu 
\eqno{(2.18)}  $$
as
$$ {{\cal L}'}_I = -Z_1 e\lambda_s^{\epsilon\over 2}  A^\mu \bar\psi \gamma_\mu \psi 
= - Z_1e \lambda_s^{\epsilon\over 2} {1\over Z_2\sqrt{Z_3}} A_{\rm b}^\mu 
\bar\psi_{\rm b} \gamma_\mu \psi_{\rm b}  $$
$$ = - e_{\rm b}  A_{\rm b}^\mu \bar\psi_{\rm b} \gamma_\mu \psi_{\rm b}
+ \textrm{finite terms} \hspace{1cm} \eqno{(2.19)} $$
where the bare charge $e_{\rm b} $ is defined as
$$ e_{\rm b}  \equiv e \lambda_s^{\epsilon\over 2} {1\over \sqrt{Z_3}}.  \eqno{(2.20)} $$
Therefore, all the infinite quantities are renormalized into the physical constants 
as well as the wave functions. 

\section{Renormalization group equation}

Now, one sees that the renormalized charge $e$ depends on the properties which 
characterize the unperturbed system.  

\subsection{Dimensional regularization}
In the dimensional regularization scheme, 
$e$ depends on the momentum scale $\lambda_s$ as
$$ e_{\rm b}  = e \lambda_s^{\epsilon\over 2}\left( 1-{e^2\over 6\pi^2\epsilon} 
\right)^{-{1\over 2}} \simeq e \lambda_s^{\epsilon\over 2} \left(1+{e^2\over 12\pi^2\epsilon} 
\right) . \eqno{(3.1)}  $$
Since the bare charge $ e_{\rm b} $ should not depend on the system, 
one finds the following RG equation 
$$\lambda_s{\partial e\over \partial \lambda_s} = {1\over 12\pi^2} e^3+ O(e^5) 
 . \eqno{(3.2)} $$
This equation can be easily solved 
for $e$.  The expression for the running 
coupling constant $ \alpha (\lambda_s) \equiv {e^2\over 4\pi} $ is given as
$$ \alpha (\lambda_s) ={\alpha (\lambda_s^0) \over 1-{2\alpha (\lambda_s^0)\over 3\pi} 
\ln \left({\lambda_s\over \lambda_s^0}\right) }   \eqno{(3.3)}  $$
where $\lambda_s^0$ denotes the renormalization point for the coupling constant. 
This is the standard procedure to obtain the behavior of the coupling constant 
as the function of the $\lambda_s$. However, the $\lambda_s$ does not appear in the Hilbert 
space of the unperturbed Hamiltonian $H_0$, and therefore one should find out 
the physical quantity corresponding to the momentum scale $\lambda_s$ itself as we discussed 
above. One can see that the only possible quantity for the $\lambda_s$ in the unperturbed 
Hilbert space should be the inverse of the box length $L$, that is
$ \lambda_s \sim {1 \over L} $ as discussed in eq.(1.4). 
In this case, one should always take the thermodynamic limit at the end of 
calculation, and therefore, one should make the limit of 
$$ \lambda_s \rightarrow \lambda_s^0 \simeq 0 \eqno{(3.4)} $$
where  $\lambda_s^0 $ corresponds to the thermodynamic limit. 

\subsection{Cutoff momentum regularization}

The RG equation can be obtained in the case of 
the cutoff momentum treatment. In this case, the bare charge $e_{\rm b} $ can be written 
$$ e_{\rm b} =e+{e^3\over 12\pi^2}\ln \left({\Lambda\over m}\right)
=e+{e^3\over 12\pi^2}\ln \left({2\pi N\over mL}\right) . \eqno{(3.5)} $$
The bare charge $e_{\rm b} $ should not depend on the box length $L$, and 
therefore one can derive the constraint equation for $e$ 
$$ L{\partial e\over \partial L}={1\over 12\pi^2} e^3 + O(e^5) .  \eqno{(3.6)}  $$
Thus, one obtains for the running coupling constant $ \alpha (L)$
$$ \alpha (L) ={\alpha (L_\infty) \over 1-{2\alpha (L_\infty)\over 3\pi} 
\ln \left({L\over L_\infty}\right) }   \eqno{(3.7)}  $$
where $ L_\infty $ denotes the value which corresponds to the thermodynamic limit, and  
$\alpha (L_\infty) $ should be fit to the observed value of the fine structure constant.  
In normal circumstances, one should always take the thermodynamic limit of 
$ L \rightarrow \infty$ in order to obtain any physical observables. 
However, in case one wishes to examine the finite size effects in the model 
field theory, then one can make use of the RG equation of 
eq.(3.7). 

\subsection{Difference in Renormalization Group Equations}

It should be interesting to note that the behaviors between eqs.(3.3) and (3.7) are 
opposite to each other. The different behavior basically originates from 
the wave functions in $D$ dimensions in the dimensional regularization scheme.  
The dimensions of the fields  $\psi$ and $A^\mu$ become in the dimensional regularization
$$ [\psi]\sim \lambda_s^{ {D-1\over 2}}, \ \ \ \  [A^\mu] \sim \lambda_s^{ {D-2\over 2}} .
  \eqno{(3.8)}  $$  
Therefore, the dimension of the interaction Lagrangian density must be modified by hand as
$$ {{\cal L}}_I =-e\lambda_s^{4-D\over 2} A^\mu \bar\psi \gamma_\mu \psi  \eqno{(3.9)}  $$
since the dimension of the Lagrangian density must be 
$ [{\cal L}] \sim \lambda_s^D $. 
The factor $\lambda_s^{4-D\over 2}$ plays an important role for the sign in front of 
RG equation, and indeed it causes the different RG equations 
from the cutoff regularization scheme. But the physical significance of the different 
RG equations between the two regularization schemes is unclear. 

\section{ Conclusions}

The concept of the renormalization is originated from the perturbative treatment 
in  quantum field theory. Since it is practically impossible to find the exact 
eigenstates of the quantized Hamiltonian in quantum field theory, it is natural 
that the theoretical framework is based on the perturbative approach. 

After the renormalization, one realizes that the renormalized charge $e$ should 
depend on the momentum scale $\lambda_s$ since the bare charge should not 
depend on the system, and this is a reasonable condition. In this case, however, 
one should understand what the $\lambda_s$ indicates in terms of physical observables. 
The Hilbert space of the unperturbed Hamiltonian is well constructed, and therefore 
one must find a quantity corresponding to the $\lambda_s$ in this Hilbert space as long as 
one employs the perturbation theory. 
The only reasonable candidate for the $\lambda_s $ must be the inverse of the box length $L$ 
as shown in eq.(1.4). Therefore, one sees that the RG equation gives the finite size 
behavior of the renormalized coupling constant $e$. 

In this sense, one should always be careful for applying the result of the RG equation to 
other physical processes. The renormalization scheme itself is perfectly well constructed, 
but the RG group equation itself cannot be more than the perturbation theory, 
and it shows how the renormalized coupling constant $e$ may respond to the change 
of the system size $L$.

\end{document}